\documentclass[%
 reprint,
superscriptaddress,
 amsmath,amssymb,
 aps,
 prl
]{revtex4-1}

\usepackage{times,color}
\usepackage{graphicx}
\usepackage{dcolumn}
\usepackage{bm}

\begin{document}

\preprint{Helical edge}

\title{Evolution of Edge states and Critical Phenomena in the Rashba Superconductor with Magnetization}

\author{Ai Yamakage}
\author{Yukio Tanaka}%
\affiliation{%
Department of Applied Physics, Nagoya University, Nagoya, 464-8603, Japan
}%

\author{Naoto Nagaosa}
\affiliation{
Department of Applied Physics, University of Tokyo, Tokyo 113-8656, Japan
}
\affiliation{
Cross Correlated Materials Research Group (CMRG), ASI, RIKEN, WAKO 351-0198, Japan
}




\date{\today}

\begin{abstract}
We study Andreev bound states (ABS) 
and resulting charge transport 
of Rashba superconductor (RSC) 
where 
two-dimensional semiconductor (2DSM) heterostructures 
is sandwiched by 
spin-singlet $s$-wave superconductor and ferromagnet insulator. 
ABS becomes a chiral Majorana edge mode 
in topological phase (TP).  
We clarify that  
two types of quantum criticality about the 
topological change of ABS 
near a quantum critical point (QCP), whether 
ABS exists at QCP or not. 
In the former type, ABS has a energy gap and 
does not cross at zero energy in non-topological phase (NTP). 
These complex properties can be detected by 
tunneling  conductance between normal metal / RSC junctions.  
\end{abstract}

\maketitle


Topological quantum phenomena 
and relevant quantum criticality have been an important concept in 
condensed matter physics \cite{Wen,Volovik}. 
Recently, stimulated by the issue of Majorana fermion in condensed matter 
physics \cite{Read,Majorana1,Majorana2,Majoranap}, 
topological quantum behavior of 
superconductivity becomes a hot topic 
\cite{Schnyder,Qi,Roy,Sato,SatoTanaka,Linder}. 
One of the most crucial point is the 
property  of the 
non-trivial edge modes  in topological phase 
where edge modes are protected by the bulk energy gap.  
\par

The edge state of superconductor 
has been known from the 
study of Andreev bound state (ABS) in unconventional 
superconductors \cite{Nagai,Hu,Review}. 
In high $T_{C}$ cuprate, dispersionless 
zero energy ABS ubiquitously 
appears \cite{Hu,Review}
due to the sign change of the pair potential on the Fermi 
surface. The zero energy state manifests itself as a 
zero bias conductance peak in tunneling spectroscopy 
\cite{TK95,Review}. 
Subsequently, the presence of 
ABS with linear dispersion has been  clarified 
in chiral $p$-wave 
superconductor \cite{Matsumoto} realized in 
Sr$_{2}$RuO$_{4}$, where time reversal symmetry 
is broken \cite{Maeno}. 
On the other hand, in the presence of 
spin-orbit(SO) coupling with time reversal symmetry, 
it has been revealed that 
spin-singlet $s$-wave pairing 
and spin-triplet $p$-wave one 
can mix each other due to the broken inversion symmetry 
\cite{Iniotakis,Eschrig,Tanaka}. 
ABS appears as a helical edge mode 
appears for 
$\Delta_{p}  > \Delta_{s} $ 
where we denote  $s$-wave 
and $p$-wave pair potentials 
as $\Delta_{s}$ and $\Delta_{p}$, respectively,  
with $\Delta_{s}>0$ and $\Delta_{p}>0$ 
\cite{Tanaka,Yip}. 
\par

The critical behavior of ABS 
has been discussed in 
spin-triplet chiral $p$-wave pairing \cite{Read}. 
By changing the chemical potential $\mu$ of 
spin-triplet chiral $p$-wave superconductor from 
positive to negative, ABS as a chiral Majorana mode disappears. 
The corresponding quantum critical point is $\mu=0$. 
Although, such a quantum phase transition can be possible 
in $\nu=5/2$ fractional quantum Hall system \cite{Read} and 
cold atom \cite{Tewari,Mizushima08}, 
it is significantly difficult to 
obtain superconducting state for negative $\mu$ 
in electronic superconductors.  \par

In all of above works,  ABS is generated from 
unconventional pairing with non-zero angular momentum.
On the other hand, in the presence of 
strong SO coupling with broken time reversal symmetry, 
chiral Majorana modes can be generated from spin-singlet 
$s$-wave pairing 
\cite{Fu1,TI1}.  
Fu and Kane have revealed the presence of chiral Majorana mode
at the boundary between ferromagnet and superconductor 
generated on the surface of topological insulator (TI). 
After that manipulating Majorana mode in TI \cite{TI1} and 
in semiconductor hetero structures based on conventional spin-singlet 
$s$-wave superconductor 
have been proposed in several contexts \cite{Satos,Sau,Alicea}. 
Sau $et$ $al.$ has proposed 
a unique Rashba superconductor where 
two-dimensional electron gas (2DEG) 
is sandwiched by conventional spin-singlet $s$-wave 
superconductor and ferromagnetic insulator \cite{Sau}. 
These systems are really promising for future application of 
quantum qubit since host superconductor is robust 
against impurity scattering. 
\par

Although there have been several theoretical 
studies about the present 
RSC \cite{DasSarma,PALee,Bena},  
the feature of the Andreev bound state (ABS) and 
its relevance to the topological quantum phase transition 
has not been revealed at all. 
It is known that ABS emerges as a  chiral Majorana edge mode in TP, 
however, the  evolution of ABS in the non-topological phase (NTP) and 
its connection to quantum  phase transition have not been 
clarified yet. To reveal these problems is 
indispensable to understand the 
tunneling spectroscopy of normal metal /RSC junction system 
and future applications of quantum device.  \par

In this Letter, we study 
energy dispersions  of 
ABS in RSC composed of 
2DEG sandwiched by spin-singlet 
$s$-wave superconductor and ferromagnetic insulator. 
It is clarified that 
there are two types of quantum criticality for ABS,
\textit{i.e.}, quantum phase transition with or without ABS corresponding to 
type I and type II, respectively.
In type I, ABS can exist even at critical point 
where bulk energy gap closes and in the NTP. 
Nonzero ABS generated in the NTP 
does not cross at zero energy. 
These features are completely different from those 
in type II where 
edge states become absent both at the critical point 
and in the NTP. 
The conventional criticality of spinless 
spin-triplet chiral $p$-wave superconductor belongs 
to type II \cite{Read,Mizushima08}. 
The conductance between normal metal / RSC junction 
shows wide variety of line shapes reflecting on these novel 
quantum criticalities. 
We also show the drastic jump of the 
conductance at critical point. \par

A Hamiltonian of Rashba superconductor with magnetization 
is given by the following form \cite{Satos,Sau,Alicea} :
\begin{align}
 H(\bm k) = H_0(\bm k) + H_{\rm R}(\bm k) + H_{\rm Z} + H_{\rm S},
 \label{Hamiltonian}
\end{align}
where kinetic energy $H_0$, Rashba spin--orbit interaction (RSOI) $H_{\rm R}$, 
Zeeman interaction $H_{\rm Z}$ by exchange field from FM insulator, 
and spin-singlet 
$s$--wave pair potential $H_{\rm S}$ induced by proximity effect are
$H_0(\bm k) = \xi_{\bm k} s_0 \tau_z,
	H_{\rm R}(\bm k) = \lambda (s_x \tau_0 k_y - s_y \tau_z k_x),
	H_{\rm Z} = V_z s_z \tau_z,
	H_{\rm S} = - \Delta s_y \tau_y,$
where $s$ and $\tau$ are Pauli matrices, $s_0$ and $\tau_0$ are $2 \times 2$ 
unit matrices, describing 
electron spin and particle--hole degrees of freedom, respectively. 
We take the explicit form of kinetic energy as 
$\xi_{\bm k} = k^2/2m - \mu$ with $\mu$ being chemical potential, 
for simplicity. 
%
%
%
\begin{figure}
\centering
\includegraphics[scale=0.9]{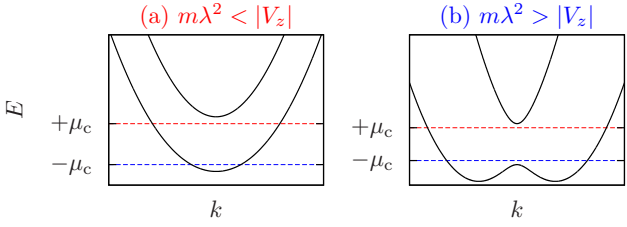}
\caption{
(color online)
Energy spectra of the normal ($\Delta=0$) states.
(a)Zeeman (Rashba spin--orbit) interaction is dominant 
with $m \lambda^2 < |V_z|$. 
(b)Rashba spin--orbit interaction is dominant 
with $m\lambda^2 > |V_z|$.
The critical value of chemical potential for the transition between topological and non--topological superconductors is given by $\pm\mu_{\rm c} = \pm\sqrt{V_z^2 - \Delta^2}$. (see discussion below eq. (\ref{chiralp}))}
\label{energy_normal}
\end{figure}
%
%
The exchange energy in a 2DEG can be tuned by changing the material of 
ferromagnetic insulator, or 
tuning the barrier thickness between the ferromagnetic insulator and the 2DEG.
%
In the normal states ($\Delta=0$),
there are two types of the energy bands as shown in Fig. \ref{energy_normal}.
For Zeeman interaction dominant case with $m\lambda^2 < |V_z|$,
there are two parabolic dispersions (Fig. \ref{energy_normal}(a)).
On the other hand, for RSOI dominant case with $m\lambda^2 > |V_z|$, 
the shape of the energy band is wine--bottle 
like (Fig. \ref{energy_normal}(b)).
As we shall see later,
the difference between these two types  
of energy bands in normal state becomes important.

The eigenvalues of the Hamiltonian for the infinite system are given by
$E_a(k_x,k_y)= \sqrt{ \eta_k + \zeta_k }$, 
$E_b(k_x,k_y) = -\sqrt{ \eta_k + \zeta_k}$, 
$E_c(k_x,k_y)= \sqrt{ \eta_k - \zeta_k }$, and 
$E_d(k_x,k_y) = -\sqrt{ \eta_k - \zeta_k}$
with
\begin{align}
\eta_k &= \xi_k^2+\lambda^2 k^2+ V_z^2 + \Delta^2,
\nonumber\\
\zeta_k &= 2 \sqrt{(\lambda^2k^2+V_z^2)\xi_k^2 + V_z^2 \Delta^2 },
\end{align}
where $k$ is defined by $k=\sqrt{k_x^2+k_y^2}$ with real $k_x$ and $k_y$ for the plane wave.
The corresponding eigenvectors $\bm u_{\alpha}(k_x,k_y)$ with $\alpha = a,b,c$, and $d$ are also obtained analytically.

Let us now consider a semi--infinite RSC in $x>0$ with flat surface at $x=0$. 
The wave function in the present system is given by
\begin{align}
 \psi_{k_y,E}(x>0) = \sum_{i=1}^{4} t_i \bm u_i(q_{i},k_y) e^{i q_i x} e^{i k_y y}.
 \label{psiT}
\end{align}
When $q_{i}$ is a real number, the corresponding wave function 
expresses  
propagating wave, \textit{i.e.}, scattering state. On the other hand, 
when $q_{i}$ is a 
complex number, it describes an evanescent wave. 
Energy $E$ and $y$--component of momentum $k_y$ are good quantum numbers.
To obtain $q_{i}$, 
we solve $k$ for fixed $E=E_{a}(k_x,k_{y})$ and $E=E_{c}(k_x,k_y)$ for $E>0$ [$E=E_{b}(k_x,k_{y})$ and $E=E_{d}(k_x,k_y)$ for $E<0$].
$q_i$ is given by $q_i=k_x$
by postulating the constraints ${\partial E_\alpha(q_i,k_y)}/{\partial q_i} > 0$  for scattering state, 
and $\mathrm{Im} q_i > 0$ for evanescent state.
Note here that, in general, $k$ and $q_{i}$ 
become  complex numbers which can be obtained 
by analytical continuation. 
The coefficient $t_i$ is determined by the confinement condition as 
$\psi_{k_y,E}(0)=0$.

\begin{figure}
\centering
\includegraphics[scale=0.4]{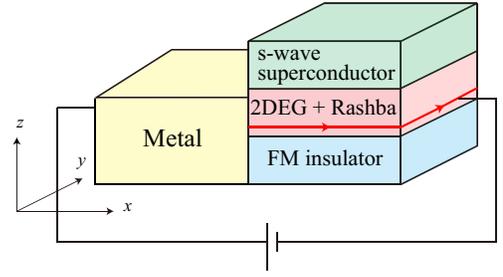}
\caption{(color online) Normal metal (N) / Rashba superconductor (RSC) junction. Andreev bound state as edge state  can exist 
denoted by the (red) arrow.}
\label{ns}
\end{figure}

\begin{figure}
\centering
\includegraphics[scale=0.74]{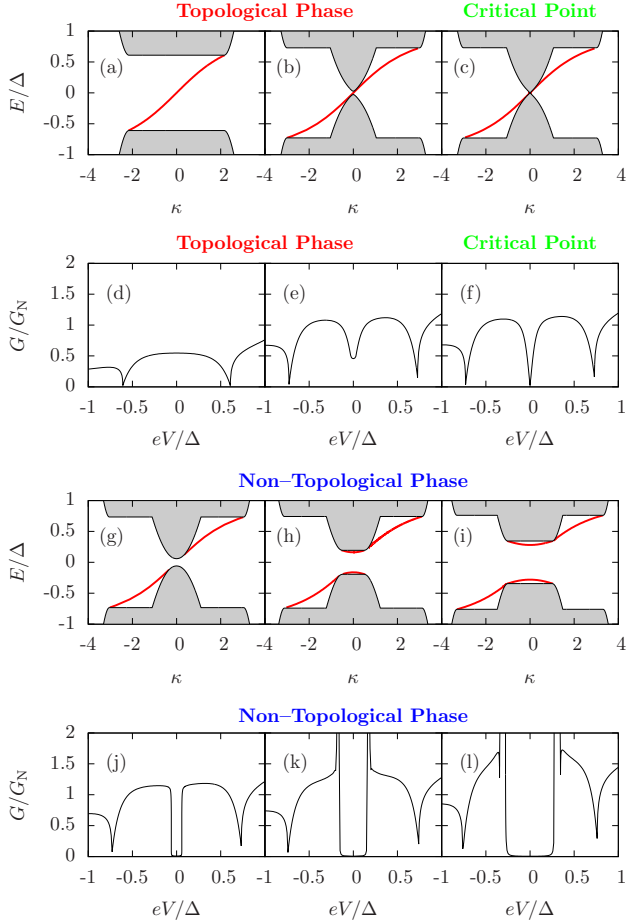}
\caption{(color online)
Energy spectra and tunneling conductances as a function of bias voltage ($eV/\Delta$) of the Rashba superconductor.
The horizontal axis denotes the normalized momentum $\kappa = k_y/\sqrt{m\Delta}$.
Zeeman interaction and Rashba spin--orbit interaction are fixed as $V_z/\Delta=2, \, m \lambda^2 /\Delta = 0.5$.
The chemical potential is set as follows. 
(a),(d):$\mu/\Delta=0$, (b),(e):$\mu/\Delta=1.7$, (c),(f):$\mu/\Delta=\sqrt 3$, (g),(j):$\mu/\Delta=1.8$, (h),(k):$\mu/\Delta=2$, (i),(l):$\mu/\Delta=2.5$.
}
\label{energy_Rashba}
\end{figure}

Tunneling conductance of normal metal (N) / RSC junction 
as shown in Fig. \ref{ns}
is calculated based on the standard way \cite{BTK,Review}. 
Suppose that
the normal metal has no spin--orbit interaction, \textit{i.e.}, the Hamiltonian reads $H_{\rm M}(\bm k) = (k^2/2m - \mu_{\rm M}) s_0 \tau_z$, where $\mu_{\rm M} = \mu - \epsilon_0$ with $\epsilon_0$ being the energy of bottom of the energy band, which is negative, and the interface 
potential is given by $H_{\rm I} = H s_0 \tau_0 \delta(x)$. 
The wave function in N is given by 
\begin{align}
 &\psi_{k_y,E,s}(x<0) 
\nonumber\\
&= \left[
	\chi_{s \rm e} e^{i k_{\rm ex} x}
	+ \sum_{s'\tau'} r_{ss'\tau'} \chi_{s' \tau'} 
	e^{ -i \tau' k_{\tau' x} x}
	\right] e^{i k_y y},
\end{align}
where the first term denotes an incident electron  with spin $s$, and $\chi_{s\tau}$ is the eigenvector of spin $s$ for electron ($\tau=+1$) or 
hole ($\tau=-1$), and $k_{\mathrm{e} x} = \sqrt{2m(\mu_{\rm M}+E)-k_{y}^{2}}$ and $k_{\mathrm{h} x} =\sqrt{2m(\mu_{\rm M}-E)-k_{y}^{2}}$ are momenta 
of reflected electron and hole, respectively.
On the other hand, the wave function in RSC ($x>0$) obeys  
the same form as in eq. (\ref{psiT}).
The boundary condition at the interface located on $x=0$ is given by the following 
two expressions \cite{yokoyama06}.
$
 \psi(-0) = \psi(+0),
 v(+0) \psi(+0) - v(-0) \psi(-0) = -i 2 H \tau_z \psi(0),
$
where velocity in $x$--direction is
$v(x) = \partial H/\partial k_x|_{k_x \to -i \partial_x}$.
Solving the above equations, we obtain reflection (transmission) coefficient $r$ ($t$).
Charge conductance $G$ normalized by its value $G_{\rm N}$ 
in the normal state ($\Delta=0$) with $V_z=0$, $\mu/m\lambda^2 = 4$, $\mu_{\rm M}/m\lambda^2 = 2 \times 10^4$, and $Z^2=mH^2/\mu_{\rm M}=10^4$, which corresponds to the case of Figs. \ref{energy_Rashba}(h) and \ref{energy_Rashba}(k) with $\Delta=0$, at zero bias voltage ($eV=0$) is given by 
\begin{align}
 G/G_{\rm N} &= \sum_s \int_{-k_{\rm F}}^{k_{\rm F}} dk_y T_s(k_y,E)
\nonumber\\ & 
\bigg /
	\sum_s \int_{-k_{\rm F}}^{k_{\rm F}} dk_y T_s(k_y,0),
\end{align}
with $T_s(k_y,E) = 2 - \sum_{s' \tau'} \tau' |r_{ss'\tau'}|^2$ and 
$\mu_{M}=2mk_{F}^{2}$. 
%
%
Hereafter, the parameters are fixed as $Z^2 = 10$, $\mu_{\rm M}/\Delta = 10^4$, and
all the conductances $G$ are normalized by the same value of $G_{\rm N}$.

\begin{figure}
\centering
\includegraphics[scale=0.7]{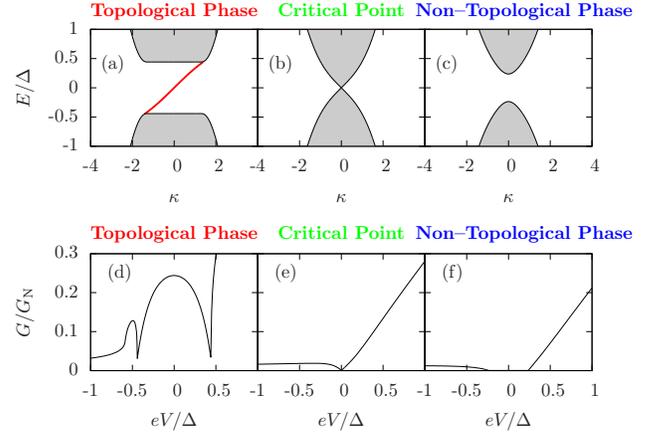}
\caption{(color online) Energy spectra (upper) and tunneling conductances (lower) of the Rashba superconductor for negative chemical potentials.
(a),(d):$\mu/\Delta=-1$, (b),(e):$\mu/\Delta=-\sqrt 3$, (c),(f):$\mu/\Delta=-2$. 
The other parameters are the same as in Fig. \ref{energy_Rashba}
}
\label{energy_chiralp}
\end{figure}

We discuss the energy spectra and the tunneling conductances, 
focusing on the difference of the criticality between two RSCs 
with different chemical potential with $\mu>0$ 
(Fig. \ref{energy_Rashba}) and $\mu < 0$ (Fig. \ref{energy_chiralp}) 
for $ \mid V_{z} \mid> m\lambda^{2}$. 

In TP (Fig. \ref{energy_Rashba}(a) and Fig. \ref{energy_chiralp}(a)), 
ABS appears as  a  chiral Majorana edge mode,  where 
$|V_z| > \sqrt{\mu^2+\Delta^2}$ is satisfied. 
Due to the presence of this 
mode, the corresponding tunneling conductance has a 
zero bias peak as shown in 
Fig. \ref{energy_Rashba}(d) and Fig. \ref{energy_chiralp}(d).
For $\mu>0$, near the QCP [Fig. \ref{energy_Rashba}(b)], 
although ABS appears as a chiral Majorana mode, 
the corresponding $G$ has a zero bias dip as shown in 
Fig. \ref{energy_Rashba}(e) 
due to the presence of a parabolic dispersion  
of bulk energy spectra near $k_y=0$.   
At QCP (Fig. \ref{energy_Rashba}(c)), it is noted that ABS remains although the bulk energy gap closes at 
$k_{y}=0$. 
This feature is quite different from $\mu<0$, where ABS is absent 
at QCP (Fig. \ref{energy_chiralp}. (b)). 
The resulting $G$ has a $V$--shaped zero energy dip both for two cases 
shown in Figs. \ref{energy_Rashba}(f) and \ref{energy_chiralp}(e). 
For $\mu > 0$, ABS still remains even in the NTP as shown in 
Figs. \ref{energy_Rashba}(g),  \ref{energy_Rashba}(h), and 
\ref{energy_Rashba}(i). 
ABS has an energy gap 
and is absent around $k_{y}=0$. 
The tunneling conductance shows a gap structure around $eV=0$ [Fig. \ref{energy_Rashba}(j)]. 
With the increase of $\mu$, $i.e.$, away from QCP, 
the additional non-zero ABS
around $k_y=0$ [Fig. \ref{energy_Rashba}(h) and \ref{energy_Rashba}(i)] 
with the almost flat dispersion are generated.
As a result, $G$ 
has two  peaks at the corresponding 
voltages inside the bulk energy gap (Fig. \ref{energy_Rashba}(k) and \ref{energy_Rashba}(l)). 
On the other hand, for $\mu<0$, ABS is absent in NTP as shown in 
Fig. \ref{energy_chiralp}(c). The resulting $G$ is almost zero inside the 
bulk energy gap (Fig. \ref{energy_chiralp}(f)). 
Based on these results, we can classify two types of criticality whether 
edge state exists at QCP or not. 
We denote former type as type I and the latter one as type II 
in the following. 
\par

\begin{figure}
\centering
\includegraphics[scale=0.74]{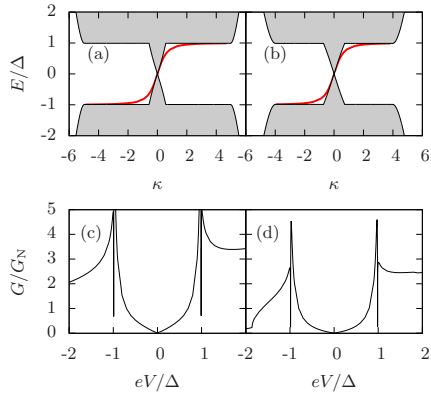}
\caption{
(color online)
Energy spectra and tunneling conductances of the Rashba superconductor for $m\lambda^2 > |V_z|$ at quantum critical point. 
(a),(c):$\mu/\Delta = \sqrt{1.25}$, (b),(d):$\mu/\Delta = - \sqrt{1.25}$.
The other parameters are taken as follows.
$m\lambda^2 /\Delta = 5, \, V_z / \Delta = 1.5$.
}
\label{energy_Rashba_dominant}
\end{figure}

We have also studied for $ |V_{z}| \leq m\lambda^{2}$. 
The energy spectra at QCP with positive $\mu$ (Fig. \ref{energy_Rashba_dominant}(a)) and negative $\mu$ (Fig. \ref{energy_Rashba_dominant}(b)) are shown.
In this case, irrespective of the value of $\mu$, 
ABS exists at QCP. Therefore, the resulting criticality is always type I. \par

Type I and II transitions can be distinguished experimentally by the line shape of $G$. 
In type I transition, line shape of $G$ becomes 
almost symmetric with respect to $eV=0$ as shown in Figs. \ref{energy_Rashba}(f), \ref{energy_Rashba_dominant}(c), 
and \ref{energy_Rashba_dominant}(d) as compared to that 
in type II as shown in Fig. \ref{energy_chiralp}(f).
Furthermore, $G$ at type I transition takes one order of magnitude larger value than that at type II, due to contribution from the edge states.

It is noted that the small value of $Z^2$ does not qualitatively change the results of the paper 
In the low transparency limit, the contribution from edge states becomes dominant for the conductance $G$, then the resulting line shape of $G$ becomes insensitive to the parameters of the normal metal, i.e., $Z^2$, $\mu_{\rm M}$, and $m$. 
In the present case, the transmission probability in the normal state ($\Delta = 0$) becomes sufficiently small with $G_{\rm N}/G_0 \sim 10^4$, where
$G_0$ denotes the maximum value of $G_N$, even for $Z^2 = 0$ since the magnitudes of Fermi momenta in left normal metal ($x < 0$) and right RSC ($x > 0$) are much different with $\mu_{\rm M}/\mu > 10^3$.

Here, we mention  the criticality of 
ABS in spinless chiral $p$-wave superconductor. 
Hamiltonian of spinless chiral $p$--wave superconductor is given by
\begin{align}
 H_{\rm p}(\bm k)
 = \begin{pmatrix}
	k^2/2m - \mu & \Delta_{\rm p} k_-
	\\
	\Delta_{\rm p} k_+ & -k^2/2m + \mu
 \end{pmatrix}.
 \label{chiralp}
\end{align}
It is known that QCP is located at $\mu=0$.  
ABS appears as a chiral Majorana mode in TP ($\mu > 0$) 
while it is absent in NTP $\mu<0$, respectively
\cite{Read}.
ABS disappears at QCP. In the light of our classification, 
quantum criticality of spineless chiral $p$-wave superconductor belongs to 
the type II. \par

To understand the difference of two types of criticality, 
we focus on the energy dispersions in the 
normal state shown in Fig. \ref{energy_normal}.
Here we introduce the critical value of transition between TP and NTP
$\pm\mu_{c}=\pm\sqrt{V_{z}^{2}-\Delta^{2}}$. 
The ABS is generated from 
$-k_{\rm F}$ to $+k_{\rm F}$, where the magnitude of 
$k_{\rm F}$ is almost the same with that of the 
large Fermi surface. 
First, we focus on the case with $m\lambda^{2}< |V_{z}|$. 
The type I 
quantum phase transition occurs at $\mu=\mu_{c}$, 
shown in Fig. \ref{energy_Rashba}. 
In this case, 
the large Fermi surface survives as shown in Fig. \ref{energy_normal}(a). 
On the other hand, as shown in Fig. \ref{energy_chiralp}, 
type II quantum phase transition occurs at $\mu=-\mu_{c}$. 
In contrast to the type I, the large Fermi surface 
vanishes in the NTP as shown in Fig. \ref{energy_normal}(a). 
For $m\lambda^{2}> |V_{z}|$, the quantum criticality 
always belongs to type I. 
Actually, as shown in  Fig. \ref{energy_normal}(b), 
the large Fermi surface survives both at $\mu=\mu_{\rm c}$ and 
$\mu=-\mu_{\rm c}$.  
For type I, the number of Fermi surfaces is 2 in NTP and 
1 in TP. 
On the other hand, for type II, 
the number of Fermi surface is 0 in NTP and 
1 in TP. 
Above rich behavior of quantum criticality in RSC originates 
from the simultaneous existence of the
Rashba spin-orbit coupling and the Zeeman interaction. 

\begin{figure}
\centering
\includegraphics[scale=0.9]{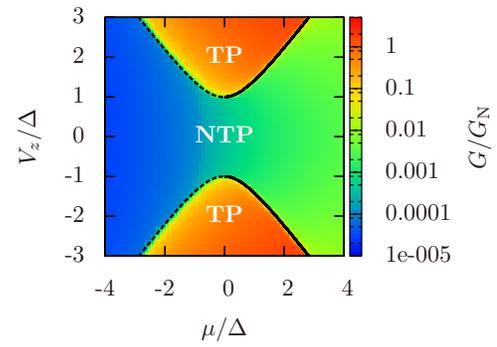}
\caption{(color online) Conductance as a function of chemical potential $\mu/\Delta$ and Zeeman interaction $V_z/\Delta$.
Rashba spin--orbit interaction is taken as $m\lambda^2/\Delta=0.5$.
The transition of type I (II) occurs at positive (negative) $\mu$.
The solid (broken) line indicates the critical line of type I (II) transition.
}
\label{gmuv}
\end{figure}

Finally, we show the zero--bias tunneling conductance of RSC as a function of $\mu$ and $V_z$ in Fig. \ref{gmuv}.
The quantum phase transition from NTP to TP occurs with tuning the parameter $V_z$ or $\mu$.
In accordance with this transition, the conductance increases by about three orders of magnitude, due to the contribution from 
zero energy ABS at $k_{y}=0$.


In this letter, 
we have calculated the energy spectrum and the tunneling conductance of 
RSC and clarified its quantum criticality.
Quantum phase transition between topological and non--topological 
superconductors has two types of criticality 
whether ABS  survives or not  at QCP. 
It is remarkable that 
ABS  can remain at QCP in RSC 
distinctly from spinless chiral $p$--wave superconductor 
which is a prototype of topological superconductor. 
This stems from the structures of Fermi surfaces 
which are spin--split by Rashba spin--orbit interaction in the normal state.
This results can provide a new perspective of quantum criticality for topological superconductors. 
We have considered only the 
spin-singlet $s$-wave superconductor. 
It is interesting to study in the case of unconventional 
superconductor where much richer quantum criticality can be expected \cite{Mizuno,Fujimoto,Schnydernew}.

This work is supported by Grant-in-Aid for Scientific Research
(Grants No. 17071007, No. 17071005, No. 19048008,
No. 19048015, No. 22103005,
No. 22340096, and No. 21244053) 
from the Ministry of Education, Culture,
Sports, Science and Technology of Japan, Strategic
International Cooperative Program (Joint Research Type)
from Japan Science and Technology Agency, and Funding
Program for World-Leading Innovative RD on Science and
Technology (FIRST Program).


\clearpage
\clearpage

\end{document}